\documentclass[floatfix,a4paper,prl,twocolumn,superscriptaddress, showpacs,amsmath,amssymb,footinbib]{revtex4-1}
\pdfoutput=1

\usepackage{natbib}
\usepackage[english]{babel}
\usepackage{letltxmacro}
\LetLtxMacro{\ORIGselectlanguage}{\selectlanguage}
\makeatletter
\DeclareRobustCommand{\selectlanguage}[1]{%
  \@ifundefined{alias@\string#1}
    {\ORIGselectlanguage{#1}}
    {\begingroup\edef\x{\endgroup
       \noexpand\ORIGselectlanguage{\@nameuse{alias@#1}}}\x}%
}
\newcommand{\definelanguagealias}[2]{%
  \@namedef{alias@#1}{#2}%
}
\makeatother
\definelanguagealias{en}{english}
\definelanguagealias{English}{english}

\usepackage{graphicx}
\usepackage[usenames,dvipsnames]{color}

\usepackage[colorlinks=true,allcolors=blue]{hyperref}

\usepackage{bbm}
\usepackage{changes}

\newcommand{\tr}[2][]{\ensuremath{\text{tr}_{#1}\left[ #2\right]}}

\newcommand{\operator}[1]{\hat{#1}}

\newcommand{\abs}[1]{\ensuremath{\left\vert #1 \right \vert}} 
\newcommand{\bra}[1]{\langle #1 |}
\newcommand{\ket}[1]{| #1 \rangle}

\def\clap#1{\hbox to 0pt{\hss#1\hss}}

\newcommand{\dfs}{\text{dfs}} 

\usepackage{layouts}

\begin{document}

\title{Strong frequency dependence of transport in the driven Fano-Anderson model}

\author{Daniel Hetterich}
\affiliation{Institute for Theoretical Physics,  University of W\"urzburg, 97074 W\"urzburg, Germany}

\author{Gabriel Schmitt}
\affiliation{Institute for Theoretical Physics,  University of W\"urzburg, 97074 W\"urzburg, Germany}

\author{Lorenzo Privitera}
\affiliation{Institute for Theoretical Physics,  University of W\"urzburg, 97074 W\"urzburg, Germany}

\author{Bj\"orn Trauzettel}
\affiliation{Institute for Theoretical Physics,  University of W\"urzburg, 97074 W\"urzburg, Germany}

\date{\today}

\begin{abstract}
We study a periodically driven  central site coupled to a disordered environment. 
In comparison to the static model, transport features are either enhanced or reduced, depending on the frequency of the drive.
We demonstrate this by analyzing the statistics of quasienergies and the logarithmic entanglement growth between bipartitions, which show similar features: 
For frequencies larger than disorder strength, localization is enhanced due to a reduced effective coupling to the central site. 
Remarkably, localization can even be increased up to almost perfect freezing at particular frequencies, at which the central site decouples due to the emergence of  ``dark Floquet states''. This high-frequency domain of our model is bounded by a critical frequency $\omega_c$, where transport increases abruptly. We demonstrate that $\omega_c$ is  determined by  one-photon resonances, which connect states across the mobility edge. This sensitive frequency dependence allows us to fine tune transport properties of the driven central site model, by unprecented precision.
 
\end{abstract}

\maketitle

\emph{Introduction.---}
 The study of the dynamics of disordered quantum systems has proven itself to be one of the most interesting phenomena in condensed matter physics. Starting with the paradigm shift given by the discovery of Anderson localization~\cite{Anderson_PR58}, 
 a great effort was put into the investigation of transport properties of disordered solid state systems, leading to many remarkable discoveries~\cite{Lee_RMP85,Phillips_ARPC93,Evers_RMP08}.
Recently, disordered systems became a platform where to address even more fundamental questions.
 In this respect, the main example is the thermalization of quantum systems. Indeed,  it came as a surprise the fact that systems undergoing many-body localization, an interacting generalization of Anderson localization, do not experience local thermalization of observables and keep memory of their initial correlations~\cite{Nandkishore_ARCMP15}. It is then natural to investigate how disordered systems behave in the presence of one of the most basic non-equilibrium protocols, namely periodic driving. In that case, novel phenomena that are completely driving-induced, such as discrete time crystals, have been proposed~\cite{Khemani2016,Else2016,Yao2017} and observed~~\cite{Zhang2017}.
 Remarkably, the degree of localization of a quantum system, can  be controlled to some extent by an external drive~\cite{Holthaus_PRL95,Holtaus_PM96,Molina_PRB06,Kitagawa_AOP12,Ponte_PRL15,Lazarides_PRL15,Roy_PRB15,Hatami_PRE16,Bairey_PRB17,Agarwal_PRB17,Bordia_NPhys17}, even across a phase transition.

   We focus on the physics properties established by a periodic drive of a disordered variant of the  Fano-Anderson model~\cite{mahan2013many}: the disordered  central site model (DCSM)~\cite{Hetterich_PRB17}. This model is interesting for fundamental reasons such as the analysis of the stability of many-body localization~\cite{Hetterich_PRB17,Ponte_PTRLA17,Hetterich_PRB18}, and for its relevance to describe central spin setups~\cite{Loss1998, Hetterich2015}. The non-interacting DCSM displays already most of the distinctive features of the interacting central spin model: It manifests for instance critical properties that are in between Anderson localized and extended phases, such as multifractal eigenstates and a logarithmic growth of entanglement entropy~\cite{Hetterich_PRB17}.
   These  properties descend purely from its star-like topology:  central site couples equally strong to each site of a localized one-dimensional chain. 
   In this work, we study the response of the DCSM to a periodic drive. Specifically, we choose to drive the central site  potential.
   
Our findings show that the driving frequency $\omega$ has an intense impact on the transport features of our model. We quantify our results by means of the growth of entanglement entropy between bipartitions and the amount of level repulsion inside the Floquet eigenvalue spectrum. Upon varying $\omega$ from infinity to energy scales that are within the bandwidth of the one-dimensional chain, transport is  reduced in a power-law fashion with respect to the undriven model. Remarkably, for some special combination of driving amplitude and frequency,  localization increases up to a point where the system becomes almost perfectly frozen. 
We demonstrate that in this case the central site, without which transport is absent, undergoes a dynamical decoupling from the chain. 
This increase of localization is contrasted by the onset of single- and multi-photon resonances as  $\omega$ is decreased, even though we find that only  those resonances that connect states across the mobility edge of the spectrum effectively delocalize the system.
Remarkably, both effects, the dynamical decoupling and the resonances can be found at the same critical frequency $\omega_c$. Hence. the system jumps abruptly from a frozen into a mobile phase of strong transport. Therefore, the driven  central site enables us to switch precisely between phases of high and low transport in disordered systems. 

\emph{Model.---} The first ingredient of the disordered central site model is a one-dimensional disordered chain, 
\begin{equation}
H_0 = \sum_{i=1}^L h_i c_i^\dagger c_i + J(c_i^\dagger c_{i+1} + \text{h.c.}),
\end{equation}
where $h_i\in[-W,W]$ are uniformly drawn random numbers. For any disorder strength $W>0$, this system experiences Anderson localization~\cite{Mott1961a}.  For simplicity, we set $J=0$ and hence assume a strong degree of localization. We now couple each site equally strong to a periodically driven central site, 
\begin{equation}
H_1(t) = A\,\text{sgn}\left[\sin (\omega t)\right] c_0^\dagger c_0 + \frac{m}{\sqrt{L}}\sum_{i=1}^L (c_i^\dagger c_0 + \text{h.c.})
\end{equation}
where $A$ and $\omega$ are the driving amplitude and frequency and $\text{sgn}$ is the signum function. Thus, the central site  couples directly to the eigenstates of $H_0$. Further, we measure energies in units of the coupling to the central site $m$, i.e. $m=1$. We then calculate numerically the stroboscopic dynamics, taking advantage of the Floquet theorem: the Floquet operator, namely the evolution operator over one period $\tau$ $ \operator{F} \equiv \hat{U}\left(\tau +t_0, t_0\right)$, gives access to the  evolution operator over $n$ periods via $\hat{U}\left(n \tau +t_0, t_0\right) = \operator{F}^n$. The Floquet operator can be written in the spectral representation as: $\operator{F} = \sum_{\alpha} \exp\left(-i E_\alpha \tau\right) \ket{\phi_\alpha(t_0)}\bra{\phi_\alpha(t_0)}$, where $E_\alpha$ are called quasienergies, while the states $\ket{\phi_\alpha}$, which are periodic in time, are called Floquet modes. The quasienergies are defined modulo $\omega = 2\pi/\tau$. In the following, we  compute the full Floquet spectrum by means of exact diagonalization, which allows us to study all dynamical and spectral properties of this model. Notably, we average over many disorder realizations until our results saturate.

\begin{figure}
\centering
\includegraphics[scale=1]{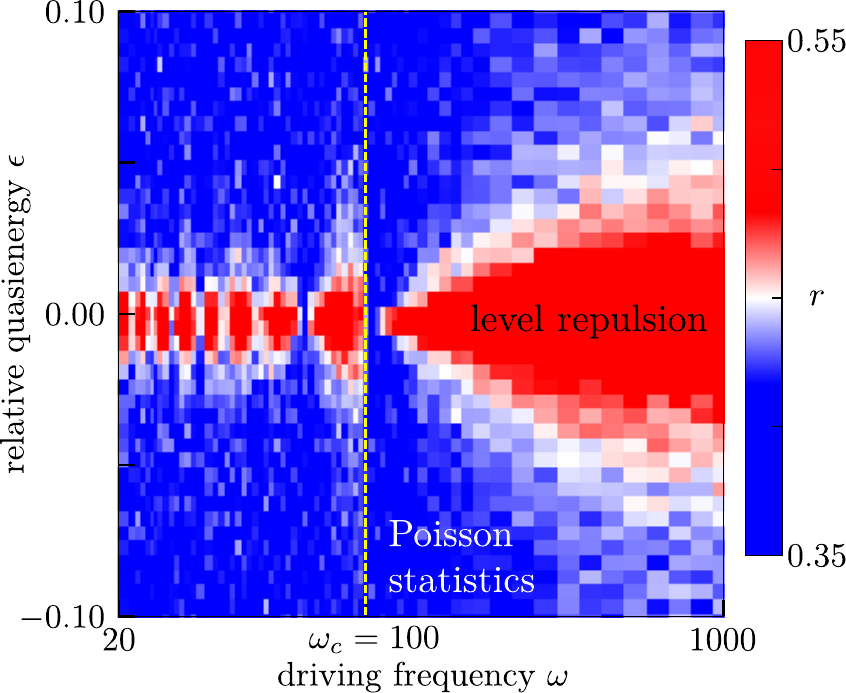}
\caption{Level statistics of quasienergies $E_i$ relative to the bandwidth, $\epsilon_i = E_i / \text{min}(W,\omega)$. For large frequencies, level repulsion is present, leading to logarithmic growth of entanglement entropy. For $\omega\gtrsim \omega_c$, localization is maximized and an almost perfect Poisson spectrum is obtained, defining our frozen regime. At $\omega=\omega_c$ (yellow dashed line) single photon resonances couple states from the center with states from the edge of the bandwidth, yielding to a drastic increase in level repulsion. Used parameters: $A=200,W=100,L=2048$.}
\label{fig1}
\end{figure}

\emph{Localization measures.---} For $A=0$, the central site model possesses a mobility edge with delocalized states in the center of the spectrum~\cite{Hetterich_PRB17}. This is because, for $A=0$, the central site mixes preferentially those Anderson states that are energetically close to it. If we now use an amplitude of $A> W$, the central site jumps between both sides of the band of $H_0$, such that it is never energetically close to the spectrum. Yet we find that only Floquet states with a quasienergy close to zero mix and yield transport features.  
We show this in Fig.~\ref{fig1}, where we illustrate the statistics of the gaps $g_i = E_{i+1}-E_i$ of the corresponding quasienergies $E_i$. In Fig.~\ref{fig1}, we employ $r=\min[g_i, g_{i+1}] / \max[g_i,g_{i+1}]$ as an indicator for level repulsion. 
We find that as $\omega$ is decreased from $\omega\to \infty$, the level repulsion region shrinks up to a point where the entire spectrum shows Poisson statistics, i.e. the system is localized. Upon  reducing the value of the driving frequency, level repulsion is suddenly reestablished  at a critical frequency $\omega_c$. 

\begin{figure}
\centering
\includegraphics[scale=1]{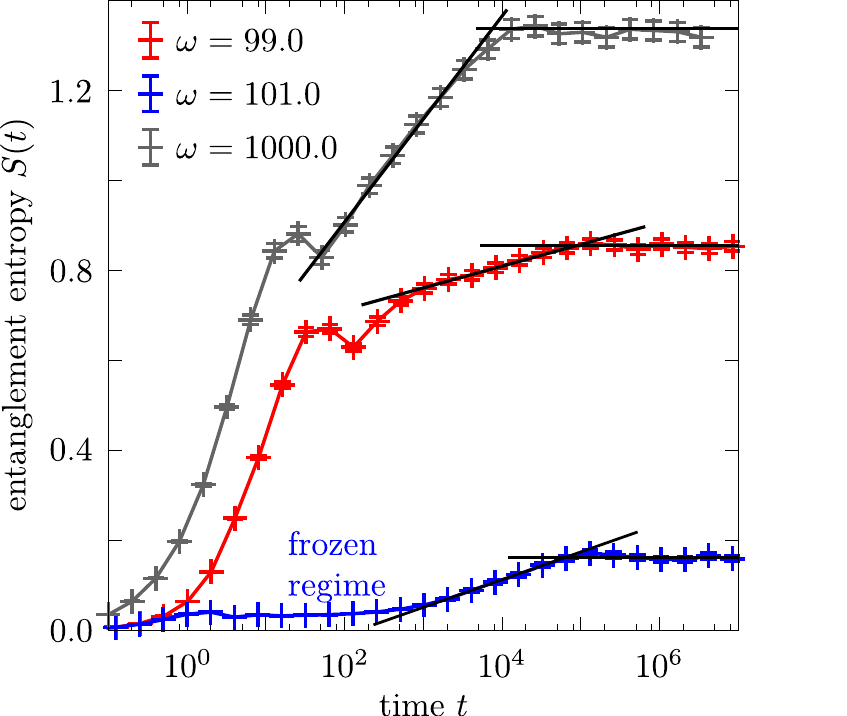}
\caption{Growth of entanglement entropy around $\omega_c=100$. For $\omega\gtrsim\omega_c$ (blue), the dynamical features of the model are frozen for many orders of magnitude in time. Crossing $\omega_c$, for $\omega<\omega_c$, entanglement growth is abruptly reactivated (red). For comparison, we show the high-frequency case $\omega=1000$, which is close to the undriven model. Solid lines indicate how we extract the slope and the saturation value, which are analyzed in the figures below. Used parameters: $A=200, W=100,L=1024$.}
\label{fig2}
\end{figure}

We demonstrate in Fig.~\ref{fig2} that this effect has also drastic impact on the dynamical features of the model. The DCSM is known to show a logarithmic growth of entanglement entropy between two contiguous halves $\mathcal{H}_A, \mathcal{H}_B$ of a bipartition $\mathcal{H} = \mathcal{H}_A\otimes  \mathcal{H}_B$ of the entire Hilbert space $\mathcal{H}$. The entanglement is thereby quantified by means of the von Neumamnn entropy 
\begin{equation}
S(t) = - \tr{\rho_A \ln \rho_A}.
\end{equation}
Here, $\rho_A = \tr[B]{\ket{\psi(t)} \bra{\psi(t)}}$ is the reduced density matrix of the time-evolved state $\ket{\psi(t)} = \exp(-i H t) \ket{\psi_0}$ and $\ket{\psi_0}$ are random many-particle eigenstates of $H_0$ at half filling. For $\omega\gtrsim \omega_c$, the growth of entanglement entropy is drastically reduced and shows an almost frozen regime before the typical logarithmic growth sets in. Due to the level repulsion that suddenly sets in for values $\omega\lesssim \omega_c$, the resulting dynamical features resemble the case of the undriven DCSM, \emph{c.f.} Fig.~\ref{fig2}.

In the following, we {demonstrate} that these findings can be explained by an interplay of several Floquet features. For $\omega \gtrsim \omega_c$, we will show that the driving effectively reduces the coupling strength of the central site with respect to the undriven model, thus enhancing  localization. This tendency is boosted by the existence of so called dark Floquet states, which give rise to the freezing effect. For $\omega = \omega_c$, single photon resonances that couple states across the mobility edge give rise to the sudden jump in transport properties. Let us develop some analytical understanding of these numerical observations.

\begin{figure}
\centering
\includegraphics[scale=1]{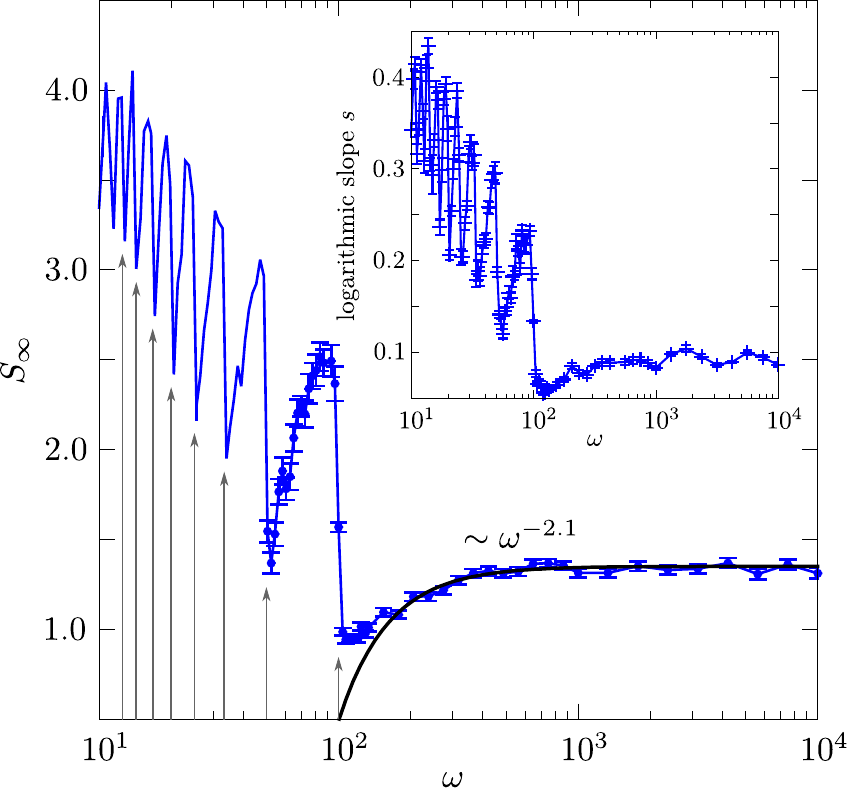}
\caption{Saturation value and logarithmic slope (inset) of the entanglement entropy. At frequencies $\omega^{\text{r}}_n = W/n=100/n$, photons resonantly couple sites from the edge of the spectrum $\pm W$ with sites from the center, causing the jumps in $S_\infty$. We indicate the first eight resonances with arrows and skip error bars for clarity in this regime. Above $\omega=100$, our data shows enhanced localization compared to the static model in agreement with our  high-frequency expansion. For the exponent of the fitting function we obtain $2.1\pm 0.3$. Used parameters: $A=100,W=100,L=1024$. }
\label{fig3}
\end{figure}

\emph{Large-frequency expansion.---} We start examining  the effect of the drive at large frequencies, where perturbative arguments can be made. Concretely, we study $\omega \gg A,W,1$. Firstly, at $\omega\to\infty$, we recover results of the undriven model, where the central site is in the center of the band. This is because at infinite frequency, any periodically driven system responds as if there is  a quench of the period-averaged Hamiltonian~\cite{Bukov_AIP15}. 
Considering effects up to $1/\omega^2$ within a Baker-Campbell-Hausdorff expansion, we find that the model can be described by simply rescaling the coupling to the central site as
\begin{equation}
m_\text{eff} = m\left( 1 - \frac{2\pi}{4} i \frac{A}{\omega} - \frac{(2\pi)^2}{24} \frac{A^2}{\omega^2}\right).
\end{equation}
The details of the expansion and the interpretation of the imaginary part of the hopping can be found in the Supplementary Material. As studied in Ref.~\cite{Hetterich_PRB17}, the dynamical properties of the system depend on second order processes in the coupling to the central site. Therefore, 
\begin{equation}
\abs{m_\text{eff}}^2 = m^2\left( 1 - \frac{(2\pi)^2}{48} \frac{A^2}{\omega^2} + \mathcal{O}\left(\frac{A^4}{\omega^4}\right) \right) \label{eq:effectivemsquare}
\end{equation}
is effectively reduced by the drive. Consequently, a periodic drive enhances the degree of localization at large values of the frequency. Indeed, if the value of $\omega$ is decreased, the picture of a resting central site in the center of the band becomes less valid, while its absence in the center establishes an energetic mismatch of the order of $A$. This perturbative regime is valid as long as $\omega$ defines the largest energy scale. 
The analytic prediction made in Eq.~\eqref{eq:effectivemsquare} is numerically testable in various ways. Among the above mentioned properties that scale as $\abs{m}^2$ in the undriven model, there is the width of the window of level repulsion inside the spectrum as well as slope and  saturation value of the entanglement entropy~\cite{Hetterich_PRB17}.
In Fig.~\ref{fig3}, we therefore study the resulting reduction of entanglement entropy (upon reducing $\omega$) until the disorder strength becomes a comparable energy scale and single photon resonances play a role. The best fit of the data agrees with Eq.~\eqref{eq:effectivemsquare} and we obtain similar results for the width of the level repulsion within Fig.~\ref{fig1} (not shown).

\emph{$n$-photon resonances.---} $n$-photon resonances change transport features due to the mixing of otherwise off-resonant degrees of freedom $E_2 - E_1 = n\omega$ at frequency $\omega$.  
Therefore, one expects the effect of single photon resonances to set in at frequencies that match the single particle spectral bandwidth $2W$. However, in the disordered central site model, the coupling is established by the central site, which is effectively at energy zero. 
A single photon resonance between the edges of the energy spectrum $\pm W$ is thus energetically suppressed by the coupling through the central site.
In contrast, at frequencies
\begin{equation}
\omega_n^{r} = W/n \;,
\end{equation}
$n$-photon resonances effectively map states from the edge of the spectrum into the center of the band, where mixing occurs. This gives rise to a drastic increase of transport, as seen in each of our figures. We note that the abrupt increase of transport is due to the uniform density of states in our model: reducing the frequency over $\omega_c:=\omega^r_1 $, an extensive amount of states is energetically mapped into the center of the spectrum, where the central site enables mixing. In Fig.~\ref{fig3}, we show the first 10 resonances $\omega_n^r$ for the saturation value $S_\infty$ and the slope $s$ of the logarithmic entanglement growth $S(t)\sim s\ln t$.

\begin{figure}
\centering
\includegraphics[scale=1]{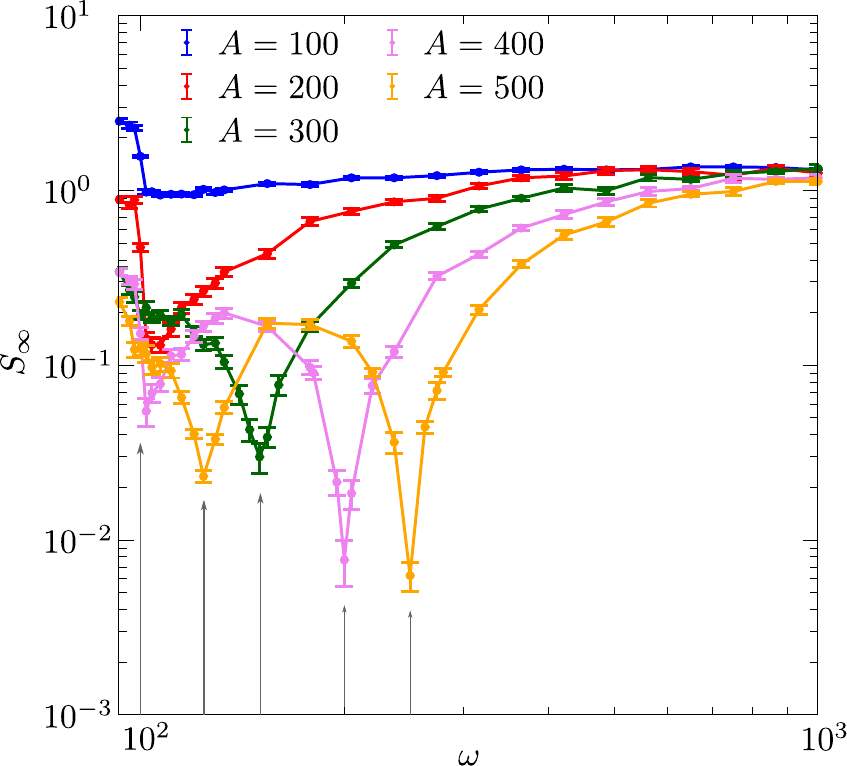}
\caption{Saturation value of the entanglement entropy in the presence of dark Floquet states. At frequencies $\omega^\dfs_n=A/(2n)$, indicated by the gray arrow, the central site decouples from the localized environment, which diminishes the entanglement entropy. Used parameters: $W=100, L=1024$.}
\label{fig4}
\end{figure}

\emph{Dark Floquet states.---} As argued before, the first photon-induced resonance is always located at $\omega_c = W$. The associated sudden discontinuity in localization measures is  particularly evident in Fig.~\ref{fig1} and~Fig.~\ref{fig2}, where freezing sets in at $\omega \gtrsim \omega_c$. In the therein presented data, we have in fact maximized the degree of localization by choosing values of $A$ that induce an additional negative resonance. As shown in Fig.~\ref{fig4}, these  resonances appear at frequencies $\omega_n^\dfs\approx A/(2n)$, a fact that cannot be explained with the perturbative approach of Eq.~\eqref{eq:effectivemsquare}, since $A\gg \omega$.  In the following we explain this freezing effect as a  sort of dynamical decoupling between the central site and its environment~\cite{Das2010}. \\
Within the first or second half of a driving period, $H(t)$ is time-independent, such that results of the undriven model can be employed. In Ref.~\cite{Hetterich_PRB17}, it was shown that the effective dynamics can be understood by virtue of the interplay between two/three sites only, including the central site. This is because all other sites are energetically off-resonant and couple only in higher orders of the weak coupling constant $m$. 

We therefore consider a single fermion on a lattice site $i$ with potential $h_i$. Its probability to be measured on the central site $c$ after time $t$ is, neglecting the rest of the system, given by
$$P_c(t) = \frac{m^2}{\Delta^2_\pm+m^2}\sin^2 \left(t\frac{1}{2}\sqrt{\Delta^2_\pm+4m^2} \right),$$ where $\Delta_\pm = \pm A - h_i$ is the potential difference between the two sites. This undriven model thus possesses an intrinsic resonance frequency $\nu_\pm = \sqrt{\Delta^2_\pm+4m^2}/2$. If the external driving frequency $\omega$ is an integer divisor of $\nu_+$, i.e. $\omega^\dfs_n = \nu_+/n = \sqrt{\Delta^2_\pm+4m^2}/(2n)$, the fermion arrives back at its initial state just at the time $T/2 = \pi/\omega_n^\dfs$, at which the central site potential changes sign in $\operator{H}_1$.

This simplistic explanation, however, does not take into account the fact that for any $h_i\neq 0$ the intrinsic frequencies differ. Their relative difference 
$$ \frac{\nu_+ - \nu_-}{\nu_+} = \frac{2 h A}{A^2 + m^2} + \mathcal{O}(h^2/A^2)$$
becomes negligible for $A\gg m, h$. Moreover, a careful series expansion of the $2\times2$ Floquet operator of this toy model reveals that   
\begin{equation}
\abs{\bra{h_i} F(\omega=\omega_1^\dfs )\ket{c} }^2 = 16 \pi^2 \frac{h_i^2m^2}{A^4} + \mathcal{O}(1/A^6),
\end{equation}
such that the Floquet eigenstates are almost perfectly localized on real space sites. Thus, the above described mechanism is robust under distinct intrinsic frequencies $\nu_\pm$ within two different parts of the driving period. 
The eigenvalue that corresponds to the Floquet eigenstate localized on the central site, obeys a complex phase of $\phi = 0 - \frac{-4i h_i m}{A^3} + \mathcal{O}(1/A^4)$, yielding a quasienergy of zero. A decoupled degree of freedom with zero quasienergy has been dubbed `dark Floquet state'~\cite{Luo_NJP14}. In our model, this dark state has approximately overlap one with the central site. Therefore, at frequencies $\omega^\dfs_n \approx A/(2n)$, transport features drop due to a dynamical decoupling of the central site. This is the main result of our work.
We remark that this suppression of transport, differs substantially from the usual coherent destruction of tunnelling, since the latter is associated with the existence of degenerate quasienergies~\cite{Grossmann_PRL91,Grossmann_EPL92}. 

\emph{Conclusions.---}
We have studied the effects of driving the central site potential in the disordered central site model.  The crucial role played by the central site in enabling transport in the system, results in an enhancement of localization as the frequency is decreased from $+\infty$. In particular, we find that for some specific values of frequency and amplitude, the central site is dynamically decoupled from the disordered chain, implying an almost perfect suppression of transport. This fact is due to the appearance of a Floquet state with zero quasienergy and overlap $\approx 1$ with the central site, analogously to what has been called dark Floquet state~\cite{Luo_NJP14}. When the frequency is sufficiently lowered, a series of multiphoton resonances across the mobility edge oppose to this trend, resulting in a significant increase of transport. Hence, the dynamical control properties of our model is exceptional. 

We thank F. Dominguez, F. Keidel and N. Traverso Ziani for discussions.
We acknowledge financial support by the DFG via Grant No. TR950/8-1, the SFB1170, and the ENB Graduate School on Topological Insulators.

%

\end{document}


\title{Supplementary Material for: \\ Strong frequency dependence of transport in the driven Fano-Anderson model}

\author{Daniel Hetterich}
\affiliation{Institute for Theoretical Physics,  University of W\"urzburg, 97074 W\"urzburg, Germany}

\author{Gabriel Schmitt}
\affiliation{Institute for Theoretical Physics,  University of W\"urzburg, 97074 W\"urzburg, Germany}

\author{Lorenzo Privitera}
\affiliation{Institute for Theoretical Physics,  University of W\"urzburg, 97074 W\"urzburg, Germany}

\author{Bj\"orn Trauzettel}
\affiliation{Institute for Theoretical Physics,  University of W\"urzburg, 97074 W\"urzburg, Germany}

\date{\today}

\maketitle

\subsection{Details on the BCH expansion}
Given the two-step form of our drive, we can derive an effective Hamiltonian in the high-frequency regime with a Baker-Campbell-Hausdorff (BCH) formula. If we choose $t_0=0$ as our initial time,  the effective (Floquet) Hamiltonian $\operator{H}_F$ is defined via
\begin{equation}
\operator{F} =  \hat{U}\left(\tau , 0\right) \equiv e^{-i \operator{H}_F \tau} = e^{-i \operator{H}_1 \frac{\tau}{2}} e^{-i \operator{H}_2 \frac{\tau}{2}} \;
\end{equation}
where $H_{1/2}$ are 
\begin{equation}
\begin{aligned}
H_{1/2} = \sum_i (\frac{m}{\sqrt{L}} (a_i^{\dagger} a_1^{\phantom{\dagger}} + a_1^{\dagger} a_i^{\phantom{\dagger}})+\epsilon_i^{\phantom{\dagger}} a_i^{\dagger} a_i^{\phantom{\dagger}}) \pm A a_1^{\dagger} a_1^{\phantom{\dagger}} \;.
\end{aligned}
\end{equation}
Using the standard BCH formula we obtain, up to second order in $1/\omega$,%
\begin{equation}
\begin{split}
H_F= \frac{H_1 + H_2}{2} +& i2\pi\frac{\lbrack H_1, H_2 \rbrack }{8\omega}  + \\ 
+ \frac{(2\pi)^2}{96\omega^2} &(- \lbrack H_1 \lbrack H_1, H_2\rbrack \rbrack + \lbrack H_2 \lbrack H_1, H_2 \rbrack \rbrack ) \;.
\end{split}
\end{equation}
The commutators are easily computed as
\begin{equation*}
\begin{aligned}
&\lbrack H_1 , H_2 \rbrack 
= - 2 A \frac{m}{\sqrt{L}} \sum_i ( a_i ^{\dagger} a_1 ^{\phantom{\dagger}} - a_1 ^{\dagger} a_i    ) \;,\\
&\lbrack H_1 \lbrack H_1, H_2\rbrack \rbrack - \lbrack H_2 \lbrack H_1, H_2 \rbrack \rbrack =
4 \frac{ A^2 m}{\sqrt{L}} (a_1^{\dagger} a_i^{\phantom{\dagger}} + \mathrm{h.c.} ) \;,
\end{aligned}
\end{equation*}
yielding the Floquet Hamiltonian
\begin{equation}
\begin{aligned}
H_F =&   \sum_i \left(\frac{m}{\sqrt{L}} \left(a_i^{\dagger} a_1^{\phantom{\dagger}} +\mathrm{h.c.}\right)+\epsilon_i^{\phantom{\dagger}} a_i^{\dagger} a_i^{\phantom{\dagger}} \right)+\\ & -\frac{2\pi}{4\omega} i A  \frac{m}{\sqrt{L}} \sum_i \left( a_i ^{\dagger} a_1 ^{\phantom{\dagger}} - \mathrm{h.c.}   \right) + \\&- \frac{ A^2 (2\pi)^2 }{24 \omega^2}\frac{m}{\sqrt{L}} \left(a_1^{\dagger} a_i^{\phantom{\dagger}} + \mathrm{h.c.}\right)  + \mathcal{O}\left(\frac{1}{\omega^3}\right).
\end{aligned}
\end{equation}
As we mention in the main text, the first order correction gives an imaginary hopping term, breaking time-reversal symmetry (TRS). However, this is only apparent, because this term depends on the initial time, while the long-time dynamics should not. Changing the initial time is equivalent to a gauge transformation~\cite{Eckardt_NJP15} on $\operator{H}_F$. Indeed, if one, for example, moves the initial time at $\tau/2$, this term changes its sign. Many quantities must be gauge independent, e.g. should not depend on the initial time of the driving. One example are the quasienergies, which we have used as a measure of localization through their gap statistics. If one is instead interested in an expansion which is gauge invariant, a tripartition of the Floquet operator must be employed~\cite{Goldman_PRX14,Eckardt_NJP15}: $\operator{F}_{t_0} = \operator{S}^{\dagger}\exp \left(-i\operator{\tilde{H}}_{F} \tau \right)\operator{S}^{\phantomsection{\dagger}}(t_0)  $. $\operator{S}$ are called kick operators, while the new Hamiltonian $\operator{\tilde{H}}_{F}$ does not depend on the initial time $t_0$ anymore. For our model, this procedure essentially consists of keeping in $\operator{\tilde{H}}_F$ only the second order real term. On the other hand, the kick operators will contain, among others, the first order term. In this way, one can separate between long time effects and intraperiod ones. For simplicity, we display in the main text the BCH result, as the localization behaviour does not change qualitatively.